\documentclass[prd,twocolumn,nofootinbib,floatfix,preprintnumbers,letterpaper]{revtex4}
\usepackage{graphicx}
\input{epsf}
\usepackage{epsf}
\usepackage{graphicx,epsfig}
\usepackage{bm}
\usepackage{latexsym,amssymb,amsmath,float,url}

\def\rd{{\rm d}}

\def\bs{\begin{split}}
\def\es{\end{split}}

\def\5{\overline 5}

\renewcommand{\(}{\left(} 
\renewcommand{\)}{\right)} 
\renewcommand{\[}{\left[} 
\renewcommand{\]}{\right]}

\newcommand{\beq}[1]{\begin{equation}\label{#1}}
\newcommand{\eeq}{\end{equation}}
\newcommand{\bea}[1]{\begin{eqnarray}\label{#1}}
\newcommand{\eea}{\end{eqnarray}}

\def\be{\begin{equation}}
\def\ee{\end{equation}}
\def\ba{\begin{eqnarray}}
\def\ea{\end{eqnarray}}

\begin{document}

\date{\today}

\title{On the dangers of using the growth equation on large scales in the Newtonian gauge.}

\author{James B. Dent and Sourish Dutta}
\affiliation{Department of Physics and Astronomy, Vanderbilt University,
Nashville, TN  ~~37235}

\begin{abstract}
We examine the accuracy of the growth equation $\ddot{\delta} + 2H\dot{\delta}  - 4\pi G\rho\delta = 0$, which is ubiquitous in the cosmological literature, in the context of the  Newtonian gauge. By comparing the growth predicted by this equation to a numerical solution of the linearized Einstein equations in the $\Lambda$CDM scenario, we show that while this equation is a reliable approximation on small scales ($k\gtrsim $h Mpc$^{-1}$), it can be disastrously inaccurate ($\sim 10^4\% $) on larger scales in this gauge. We propose a modified version of the growth equation for the Newtonian gauge, which while preserving the simplicity of the original equation, provides considerably more accurate results. We examine the implications of the failure of the growth equation on a few recent studies, aimed at discriminating general relativity from modified gravity, which use this equation as a starting point. We show that while the results of these studies are valid on small scales, they are not reliable on large scales or high redshifts, if one works in the Newtonian gauge. Finally, we discuss the growth equation in the synchronous gauge and show that the corrections to the Poisson equation are exactly equivalent to the difference between the overdensities in the synchronous and Newtonian gauges.
\end{abstract}

\maketitle

\section{Introduction}
\label{sec:introduction}
The past few decades have witnessed a remarkable improvement in the level of precision in cosmological observations.  From measurements of the temperature fluctuations of the CMB, along with the total energy budget and flatness of the universe, to distance measurements and determinations of the Hubble parameter, errors on the order of a few percent are now commonplace.  Within an arena of such precision, theorists must be especially cautious of deeply embedded approximations within their calculations. These approximations may be harmless under a given set of assumptions (for example, small scales and small redshifts), but dangerous if implemented in situations where those assumptions are not applicable (e.g. future probes on large scales and high redshifts). 

With this in mind, in the present work we examine the growth equation in the Newtonian gauge (we discuss the relation between the conformal Newtonian gauge and the synchronous gauge in Sec.\ref{AppI}), the derivation of which involves a number of approximations.  The most important of these, as we demonstrate,  is the Poisson equation which is used to link the matter perturbations (and subsequently their growth) to the metric perturbations.  We show that in linearized general relativity, the Poisson equation follows from one of the Einstein constraint equations, in which two terms have been discarded. While on small scales these terms can be safely neglected, we show that on large scales or large redshift, at least one of them can be on the same order as the perturbative variables, and its absence can introduce significant errors in the growth equation.  That these terms may be dominant on large, horizon size scales has of course been known for a long time.  What has not been appreciated is that on sub-horizon scales these terms may generate contributions which are on the order of experimental precision.  We estimate the error arising from the growth equation on various scales and redshifts, and show that the error can be significant enough in the context of current and proposed experimental limits that caution should be exercised in its use in calculations.  

Being aware of this error, and thus avoiding it, is important for many models which attempt to break the degeneracy between modified gravity (MG) and dynamcial dark energy (DDE) (see for example \cite{zhang,acquaviva,Linder:2005in,lindercahn,polarski07,dore08}).  These MG studies attempt to discern a deviation from general relativity (GR) in observations such as gravitational lensing, large scale structure growth, and the Integrated Sachs-Wolfe (ISW) effect.  The Newtonian gauge is used in these investigations due to the metric perturbations being the gravitational potential(s) (these two perturbations are equivalent in the case of no anisotropic stress, and the difference between the two is parameterized in MG by what is called the gravitational slip).  For example in lensing studies, it is the gradient of these potentials which is relevant (which is just given by the Poisson equation in the case of no anisotropic stress), in the ISW it is the time variation of the potentials which enter the calculation, and in structure growth it is the Poisson equation which is used to derive the growth equation.  Working in the Newtonian gauge then, the Poisson equation arises and plays a central role, and is used throughout the MG literature.
    
In this work we will take a more detailed look at a few of these recent investigations of MG, as well as proposed measurements at large redshift.  As stated above, these models, are particularly sensitive to any adjustments to the Poisson equation as they introduce modifications to the Poisson equation in order to investigate non-standard cosmology.  The fact that modifications to GR are proposed to operate at large scales (due to the degree of accuracy to which GR has been tested at small scales) is also an indication that one should be mindful that one is not mistaking a signal for MG when in fact it is a sign that one's approximation is becoming less accurate at the scale of interest.  In other words, GR may be erroneously thought to break down when it is the approximation used, which discards terms inherent in GR, that is the culprit.

The outline of our paper is as follows. Except in the last section, our entire work is in the Newtonian gauge. In Sec.\ref{derivation} we review the derivation of the growth equation and the Poisson equation paying attention to the assumptions and approximations that enter the calculations. In  Sec.\ref{testing and improving} we  compare the growth predictions from the growth equation to a direct numerical integration of the linearized Einstein equations in the $\Lambda$CDM scenario, and demonstrate that the chief source of error in the growth equation arises from the Poisson equation.  Sec.\ref{Probing beyond-Einstein physics} studies the implications of using the growth equation in recent efforts to discriminate general relativity from modified gravity theories.    Sec.\ref{AppI} deals with relating the growth equation and matter perturbations in the synchronous and Newtonian gauges. Conclusions are found in Sec.\ref{conclusions}.

\section{Linearized Einstein gravity: the Poisson and the growth equations}
\label{derivation}

In this section we review the derivation of the growth equation, highlighting the assumptions and approximations that go into the derivation. We work in ordinary general relativity (GR) with a general stress-energy tensor (this can include DDE). We use the notation of \cite{copeland,Hwangtachyon,Hwang05,ddw}.  The perturbed metric is of the form 
\begin{eqnarray}
\hspace*{-0.2em}\rd s^2 &=& - (1+2A)\rd t^2 +
2a\partial_iB \rd x^i\rd t
\nonumber\\
\hspace*{-0.2em}&& +a^2\left[ (1+2\psi)\delta_{ij} + 2\partial_{ij}E 
\right] \rd x^i \rd x^j\,,
\end{eqnarray}
where $A$,$B$,$\psi$ and $E$ represent metric perturbations and $a$ represents the cosmic scale factor. 

Working in Newtonian gauge (we are working with coordinate time, and therefore we are not working in the ``conformal'' Newtonian gauge, as we do in Sec.\ref{AppI} when comparing with the synchronous gauge formulation), which corresponds to a transformation to a frame such that $B=E=0$, the gauge-invariant variables characterizing the metric perturbations become:
\begin{eqnarray}
\label{defPhi2}
\Phi &\equiv& A - \frac{{\rm d}}
{{\rm d}t} \left[ a^2(\dot{E}+B/a)\right] \rightarrow A\,,\\
\label{defPsi2}
\Psi &\equiv& -\psi + a^2 H (\dot{E}+B/a) \rightarrow -\psi\,.
\end{eqnarray}
The energy-momentum tensor can be decomposed as 
\begin{eqnarray}
& & T_0^0=-(\rho+\delta \rho)\,,\quad 
T^0_{\alpha}=-(\rho+p) v_{,\alpha}\,, \nonumber \\
& & T^{\alpha}_{\beta}=(p+\delta p) 
\delta^{\alpha}_{\beta}+\Pi^{\alpha}_{\beta}\,,
\end{eqnarray}
where $\Pi^{\alpha}_{\beta}$ is a tracefree anisotropic stress, and $p$, and $\rho$  denote the pressure and  density of the cosmic fluid respectively. For the moment we take the anisotropic stress to be zero.

The perturbed Einstein equations yield, at linear order,
\begin{eqnarray}
\label{pereq1}
-\Phi+\Psi&=&0\\
\label{pereq2}
 -\frac{\Delta}{a^2}\Phi+3H^2\Phi + 3H\dot{\Phi}&=&-4\pi G \delta \rho \\
\label{pereq3}
H\Phi+\dot{\Phi}&=&4\pi G a(\rho+p)v\\
\label{pereq4}
3\ddot{\Phi} + 9H\dot{\Phi}\qquad\qquad\qquad&& \nonumber\\
  +(6\dot{H} + 6H^2 +\frac{\Delta}{a^2})\Phi&=&4\pi G (\delta \rho+3\delta p) \\
\label{pereq5}
\delta \dot{\rho}+3H (\delta \rho+\delta p)&=&(\rho+p) \left(3\dot{\Phi}\right.\nonumber\\
&&+\left.\frac{\Delta}{a}v\right) \\
\label{pereq6}
\frac{[a^4(\rho+p)v]^{\bullet}}{a^4(\rho+p)}&=&\frac{1}{a} \left( \Phi+\frac{\delta p}{\rho+p}
\right)
\end{eqnarray}
where a dot, bold or otherwise, denotes a derivative with respect to coordinate time $t$, 
$H \equiv \dot{a}/a$, and any quantity preceded by $\delta$ denotes a perturbation in that quantity.  These equations determine the two metric perturbations $\Phi$ and $\Psi$, along with the velocity potential $v$ and the matter perturbation $\delta\rho$ which is conventionally re-written as $\delta \equiv \delta\rho/\rho$.

We have used the relation between $\Phi$ and $\Psi$ in Eq.~(\ref{pereq1}) in the subsequent equations.  This relation simply reflects our assumption of no anisotropic stress.   From this point on we use the Newtonian potential $\Phi$ to characterize the metric perturbation.

We now switch to Fourier space, an extremely convenient transformation in the theory of linear perturbations. In the  Fourier-transformed $k$-space, all perturbative variables (e.g. $\delta\rho$) are replaced by their Fourier transforms (e.g. $\delta\rho_k$). For convenience, we suppress the k-subscripts and in what follows (unless otherwise mentioned), all perturbative variables can be assumed to be in Fourier space. A subscript $0$ denotes the present (zero redshift) value of that quantity. 

Eq.~(\ref{pereq2}) allows us to express the energy overdensity $\delta$ in terms of the potential $\Phi$ and the background variables as follows:
\begin{eqnarray}
\label{delta}
-4\pi G\delta\rho = \frac{k^2}{a^2}\Phi +3H^2\Phi +3H\dot{\Phi}
\end{eqnarray}
One can see from this equation that on small scales where $k^2/a^2 \gg H^2$ the first term on the right hand side dominates. If one also makes the assumption that the gravitational potential $\Phi$ is slowly varying (as in the case of matter domination), then one recovers the familiar Poisson equation
\begin{eqnarray}
\label{poisson}
-4\pi G\delta\rho = \frac{k^2}{a^2}\Phi
\end{eqnarray}

These approximations feed into the derivation of an equation governing the growth of structure in linearized gravity, commonly called the ``growth equation'', which we now derive.

Starting from Eqns.(\ref{pereq5}) and (\ref{pereq6}) we find
\ba
\label{delta-dot}
\dot{\delta}&=&-3H\Theta +3(1+w)\dot{\Phi}-(1+w)\frac{k^2}{a}v\\
\label{v-dot}
\dot{v}&=&-vH\(1-3w\)-\frac{\dot{w}}{1+w}v\nonumber\\
&&\quad+\frac{1}{a}\[\Phi+\frac{w}{1+w}\delta+\frac{\Theta}{1+w}\]
\ea
where $w$ is the EoS function defined in terms of pressure and energy density as $w \equiv p/\rho$, $\Theta$ is the combination $(\delta p/\delta\rho - w)\delta$ \cite{ddw}. It can be verified that these equations are identical to Eq.~30 in~\cite{mabertschinger}.

The equation which determines the growth of the matter perturbation is found to be
\be
\label{deltadoubledot}
\begin{split}
\ddot{\delta} + \dot{\delta}(2-3w)H + \frac{k^2}{a}w\delta + \frac{k^2}{a^2}(1+w)\Phi = \\ 
 3(1+w)[\ddot{\Phi} + \dot{\Phi}(2-3w)H] + 3\dot{w}\dot{\Phi} \\
 -3H\dot{\Theta}+ [ -\frac{k^2}{a^2}+\frac{3H^2}{2}(1 + 9w) ]\Theta 
\end{split}
\ee
(one way to see this is by substituting $v$ from Eq.~(\ref{pereq3}) and $\delta$ from Eq.~(\ref{delta}) into Eq.~(\ref{pereq6}), and using the time derivative of Eq.~(\ref{pereq5})).

To get from Eq.~\ref{deltadoubledot}, which is exactly true in linear order, to the growth equation one must make several assumptions.  First we assume the metric perturbation $\Phi$ is slowly varying ad discard all its derivatives. Next, we take $\Theta\simeq0$, $\dot{\Theta}=0$, which set the effective sound speed of the cosmic fluid $\delta p/\delta\rho$ approximately equal to its equation of state $w$, and make both of them constant in time. We then take and $w\simeq 0$. These approximations are perfectly reasonable in a universe heavily dominated by pressureless perfect fluid matter, and they lead to the following much simplified equation:

\ba
\ddot{\delta} + 2H\dot{\delta} + \frac{k^2}{a^2}\Phi  = 0
\ea
Finally using the approximation Eq.(\ref{poisson}) we recover the familiar growth equation:

\ba
\label{growth}
\ddot{\delta} + 2H\dot{\delta}  - 4\pi G\rho\delta = 0 \,.
\ea

The growth equation is ubiquitous in cosmology and is used in many different forms, usually as either a first or second order differential equation in a certain growth variable, which is a function of $\delta$. We list three versions of this equation, which will be relevant to this work, below. 

One choice of the growth factor is $G\equiv d\textnormal{ln}(\delta/a)/d\textnormal{ln}a$ which leads to the equation:
\ba
\label{Ggrowth}\nonumber
\frac{dG}{d\textnormal{ln}a} &+& \left(4 + \frac{1}{2}\frac{d\textnormal{ln} H^2}{d \textnormal{ln} a}\right)G + G^2\\ &+& 3 + \frac{1}{2}\frac{d \textnormal{ln} H^2}{d \textnormal{ln} a} - \frac{3}{2}\Omega_{m}(a) = 0
\ea

A slightly different choice of growth factor is $f \equiv d\textnormal{ln}\delta/d\textnormal{ln}a$ leading to the equation
\ba
\label{fgrowth}
\frac{df}{d\textnormal{ln}a} + f^2 + \frac{1}{2}\left(1 - \frac{d\textnormal{ln}\Omega_m}{d\textnormal{ln}a}\right)f - \frac{3}{2}G_N\Omega_m = 0
\ea
Clearly $f=G+1$. 

A third choice of growth factor $D \equiv \delta_{m}(k,z)/\delta_m(k,z\rightarrow \infty)$ yields the equation:
\ba
\label{Dgrowth}
\frac{d^{2} \ln D}{d\(\ln a\)^2} + \(2 + \frac{1}{H}\frac{d H}{d\ln a}\)\frac{d\ln D}{d\ln a} - \frac{4\pi G}{H^2}\rho D = 0
\ea

The question is then: how accurate is the growth equation upon use of these approximations?  Certainly in the matter dominated era ($z \gtrsim 1$) and on small scales ($k\gg.001$h Mpc$^{-1}$) these approximations appear justified.  However, in this era of precision cosmology one must be careful to avoid throwing away terms which may account for measurable deviations.  We now address the size and seriousness of such errors.

\section{Testing and improving on the growth equation}
\label{testing and improving}

In this section, we discuss in detail the errors that arise in the growth equation from assuming the Poisson equation to be true in the Newtonian gauge. We assume a model consisting of perfect fluid dark matter and a cosmological constant dark energy.  We test the growth equation Eq.~\eqref{growth}  by comparing it to a numerical integration of the full set of linearized Einstein equations governing the growth of perturbations in this $\Lambda$CDM Universe. We also ignore radiation in the calculation presented here, but we have verified that including radiation does not change our conclusions. 

Note that a more complicated model including baryons and neutrinos can introduce further sources of divergence between the growth equation and the true growth in the Newtonian gauge. Baryons have been shown to significantly affect the growth of the overdensity (contributing an error of 10\% on scales below 10 Mpc), and on large scales the neutrino anisotropic stress cannot be ignored. Details on both these effects can be found in \cite{Green:2005kf}. Stochastic corrections to $\delta$ as a result of random forces (possibly arising from the ``graininess'' of the underlying system of particles) was explored in \cite{stochastic} where it was shown that these random forces can lead to significant deviations from the nonstochastic solution at late times. However, we ignore these effects here, and focus only on the errors arising from relativistic corrections to the Poisson equation. 

The zero-th order Einstein and Euler equations describing the evolution of the background of this system are the following:

\ba
2\dot{H}+3H^{2}&=&8\pi G\rho_{\Lambda}\\
\dot{\rho}&=&-3H\rho
\ea
subject to the constraint
\be
3H^{2}=\left(\frac{\dot{a}}{a}\right)^{2}=8\pi G\(\rho+\rho_{\Lambda}\)\\
\ee

Here $\rho_{\Lambda}$ is the energy density of the cosmological constant. These equations can be solved analytically  \cite{gron} to yield:
\ba
\label{gron1}a(t)&=&\left[\frac{\Omega_m}{1-\Omega_m}\right]^{1/3}\sinh^{2/3}\(\frac{t}{t_{\Lambda}}\)\\
\label{gron2}H(t)&=&\frac{2}{3t_{\Lambda}}\coth\(\frac{t}{t_{\Lambda}}\) \\
\rho(t)&=&\rho_{\Lambda}\sinh^{-2}\(\frac{t}{t_{\Lambda}}\)
\ea
where $t_{\Lambda}=2/\sqrt{24\pi G\rho_\Lambda}$

The first-order equations governing the evolution of perturbations can be easily derived. Using $v_f=-va$ we find from equations \eqref{pereq4}, \eqref{delta-dot} and \eqref{v-dot}:

\ba
\label{Lper1}\ddot{\Phi}&=&-4H\dot{\Phi}-8\pi G \rho_{\Lambda}\Phi\\
\label{Lper2}\dot{\delta}&=&3\dot{\Phi}+\frac{k^2}{a^2}v_{f}\\
\label{Lper3}\dot{v}_{f}&=&-\Phi
\ea
Equations \eqref{pereq2} and \eqref{pereq3} yield the constraints:
\ba
\label{Lcons1}3H\(H\Phi+\dot{\Phi}\)+\frac{k^2}{a^2}\Phi&=&-4\pi G\delta\rho\\
\label{Lcons2}\(H\Phi+\dot{\Phi}\)&=&-4\pi G\rho v_{f}
\ea

An alternative method of deriving Eqs.(\eqref{Lper1}-\eqref{Lcons2}) is to start with the zero-th and first order Einstein equations for a Universe consisting of matter and scalar field dark energy (these can be found in, for instance \cite{liddle} in the  Newtonian gauge and \cite{duttamaor} in the synchronous gauge), and then to take the limit as the scalar field tends to a cosmological constant, that is, the potential $V\(\phi\)=\rho_\Lambda={\rm constant}$ and $\dot{\phi}\(t=0\)=0$.

For the numerical integration, we set initial conditions at a redshift $z=1100$ where we set $\delta=10^{-5}$ and $\dot{\Phi}=0$. Equations Eq.\eqref{Lcons1} and Eq.\eqref{Lcons2} are used as independent checks on the accuracy of our numerics. We call the growth predicted by our numerical evolution the ``true growth''.  A simple mathematica notebook which performs the above numerical integration can be found at \cite{website}. 

Henceforth, we denote the true growth by the variable $\delta$ and that predicted by the growth equation \eqref{growth} by $\delta_g$. We express the percentage departure of $\delta_g$ from $\delta$ by the quantity
\be
\label{Delta}
\Delta\equiv\frac{\(\delta_g-\delta\)}{\delta}. 
\ee

\begin{figure}
	\epsfig{file=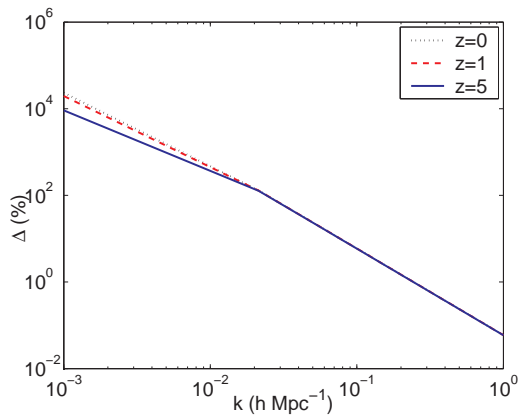,height=55mm}
	\caption
	{\label{Delta_vs_k}
The (percentage) error  $\Delta$ in the growth ($\delta_{gi}$) predicted by the usual growth equation Eq.~\eqref{growth} as a function of scale, for three different redshifts }
\end{figure}

\begin{figure}
	\epsfig{file=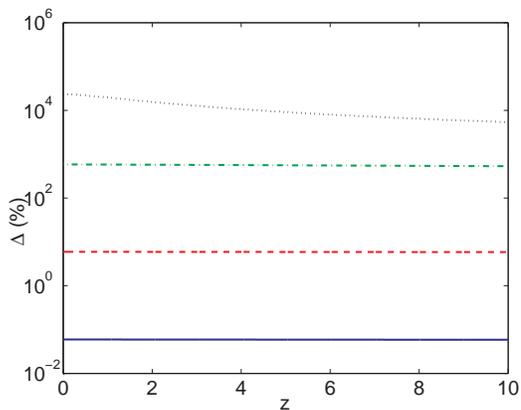,height=55mm}
	\caption
	{\label{Delta_vs_z}
The (percentage) error  $\Delta$ in the growth ($\delta_{g}$) predicted by the usual growth equation Eq.~\eqref{growth} as a function of redshift, for four different scales. From the top down, the curves are for $k=.001$h Mpc$^{-1}$, $k=.01$h Mpc$^{-1}$, $k=.1$h Mpc$^{-1}$ and $k=$h Mpc$^{-1}$ respectively.   }
\end{figure}

Fig. \ref{Delta_vs_k} shows $\Delta$ as a function of scale, and we can see that the growth equation fails spectacularly as one approaches the horizon scale. This is also reflected in Fig. \ref{Delta_vs_z}, which shows $\Delta$ as a function of redshift for different scales.  The failure at horizon scales is not a surprise of course.  What \emph{is} of interest is that there are significant deviations arising at scales within reach of experiment, and at levels that could be misinterpreted as a signal of the breakdown of GR if one were unaware that the Poisson equation approximation itself is breaking down at these scales.

Aside from $\Lambda$CDM, we have also verified the failure of the growth equation in a model where the dark energy is a scalar field with a quadratic potential (with mass comparable to the present value of the Hubble parameter), e.g. the scenario considered in detail in \cite{duttamaor}.

Having established the failure of the growth equation on large scales, we now proceed to investigate the cause for this failure and propose an improved version of this equation. On large scales, the most obviously suspect step in the derivation of Eq.~\eqref{growth} is the replacement of Eq.~\eqref{delta} by the Poisson equation, or in other words, the neglect of the term $3H(H\Phi+\dot{\Phi})$ in comparison to the term $\(k^2/a^2\)\Phi$. While this is justified on small scales $k^2\gg a^2 H^2$, it is not justified on scales close to the horizon, where the terms $3H^2\Phi$ and $\(k^2/a^2\)\Phi$ are on the same order. The error introduced in neglecting a term of the same perturbative order is rapidly magnified in the process of integrating over the age of the Universe, leading to the gigantic deviation from the true growth as shown in Figs.\ref{Delta_vs_k}-\ref{Delta_vs_z}.

A smaller error is introduced in ignoring the velocity of the metric perturbation $\Phi$. However, we have verified that this does not lead to a significant error except on horizon scales.

\begin{figure}
	\epsfig{file=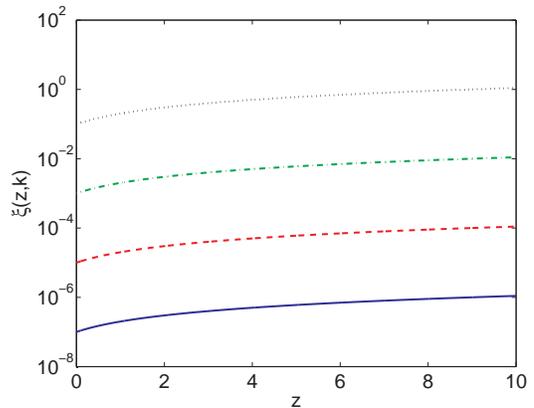,height=55mm}
	\caption
	{\label{zeta}
The correction term $\xi$ as a function of redshift, for different scales according to Eq.(\ref{xion}) as function of redshift, for four different scales. From top down, the curves are for $k=.001$h Mpc$^{-1}$, $k=.01$h Mpc$^{-1}$, $k=.1$h Mpc$^{-1}$ and $k=h$ Mpc$^{-1}$ respectively.}
\end{figure}

To understand the magnitude of the error introduced by ignoring the term proportional to $\Phi$, let $\xi\(z,k\)$  denote the ratio of the $3H^2\Phi$ term to the $\(k^2/a^2\)\Phi$ term, i.e.,

\ba
\label{xi2}
\xi = 3a^{2}H^{2}/k^2
\ea
We can estimate the size of this term (in the matter dominated regime, for simplicity)  by using 
\ba
H^2 = H_0^2 (1 + z)^3\Omega_{m0}\,\,\,;\,\,\,a = \frac{a_0}{1+z}
\ea
where the subscript zero denotes today's value, and $a_0 = 1$. For convenience, we express $k$ as
\ba
k = 10^{-n}\frac{h}{Mpc}
\ea
such that $n$ now denotes the scale. Finally, the present value of the Hubble parameter $H_0$ is expressed the usual way, as 
\ba
H_0 = 100\frac{h}{Mpc}\frac{km}{s}
\ea
Plugging the above into Eq.(\ref{xi2}), using $\Omega_{m0} = .3$ (and of course restoring the correct units by dividing by the speed of light squared),  we obtain the following simple formula for $\xi$ which shows its dependence on both physical scale and redshift:
\ba
\label{xion}
\xi = (1+z)10^{2n-7}
\ea
 In Fig.\ref{zeta} we plot $\xi$ as a function of $z$ at various scales.  One can see that the discarded term ($3H^2\Phi$) can be a substantial, non-negligible fraction of the retained term $\(k^2/a^2\)\Phi$ as $n$ becomes larger than 2 (i.e., $k\leq 10^{-2}$h Mpc$^{-1}$), or at large redshifts. At $n \approx 3$, even at small redshifts this term will be $\gtrsim \mathcal{O}(10)\%$. As mentioned before, the cumulative effect of integrating over the history of the Universe causes the error to magnify rapidly.

Based on the above considerations, we propose the following modification to the growth equation
\be
\label{growthbetter}
\ddot{\delta} + 2H\dot{\delta}  - \frac{4\pi G\rho}{1 + \xi}\delta = 0
\ee

\begin{figure}
	\epsfig{file=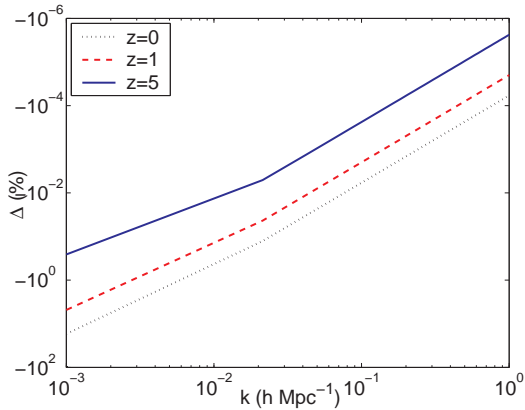,height=55mm}
	\caption
	{\label{Delta_vs_k_improved}
The (percentage) error  $\Delta$ in the growth ($\delta_{gi}$) predicted by the improved growth equation  from Eq.~\eqref{growthbetter} as a function of scale, for three different redshifts }
\end{figure}

\begin{figure}
	\epsfig{file=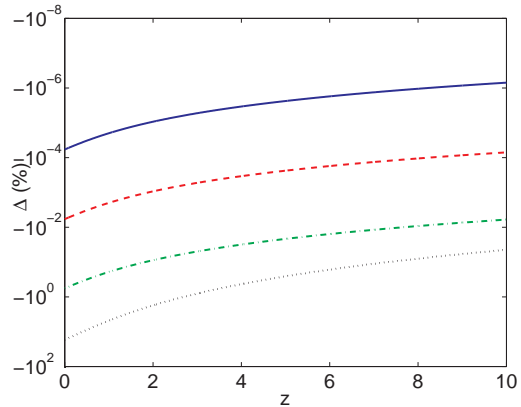,height=55mm}
	\caption
	{\label{Delta_vs_z_improved}
The (percentage) error  $\Delta$ in the growth ($\delta_{gi}$)  predicted by the improved growth equation Eq.~\eqref{growthbetter} as a function of redshift, for four different scales. From the bottom up, the curves are for $k=.001$h Mpc$^{-1}$, $k=.01$h Mpc$^{-1}$, $k=.1$h Mpc$^{-1}$ and $k=h$ Mpc$^{-1}$ respectively.  }
\end{figure}

Figs.\ref{Delta_vs_k_improved}-\ref{Delta_vs_z_improved}, which show the $\Delta$ arising from the improved growth equation (let us call this $\delta_{gi}$), verify that Eq.(\ref{growthbetter}) provides a far better approximation to the true growth than the usual growth equation. Even on scales of the size of .01h Mpc$^{-1}$ , the error is on the order of a percent for all redshifts. For scales on the order of the horizon, the error is somewhat large for low redshifts, presumably as a result of the dark energy dominance which causes the $\dot{\Phi}$ term in Eq.~\ref{delta} to become significant.

The cumulative effect of integrating a small error over a large time period can be demonstrated by considering the difference between the usual growth equation Eq.(\ref{growth}) and the improved growth equation Eq.(\ref{growthbetter}). Taking the growth predicted by the improved growth equation $\delta_{gi}$ as a proxy for the true growth (according to Fig.\ref{Delta_vs_k_improved}, the difference between $\delta$ and $\delta_{gi}$ is negligible up to scales of $.01h$ Mpc$^{-1}$), one can define the error variable in terms of $\delta_{gi}$ (rather than $\delta$ as in Eq.~\eqref{Delta}):
\be
\tilde{\Delta}\equiv\frac{\(\delta_g-\delta_{gi}\)}{\delta_{gi}}. 
\ee 
where $\delta_g$ and $\delta_{gi}$ evolve according to Eqs.(\ref{growth}) and (\ref{growthbetter}) respectively. The growth variables defined in this section are listed for convenience in Table ~\ref{deltas}.

One can now construct a differential equation for the evolution of $\tilde{\Delta}$ for a given $k$ (under the assumption of matter domination):
\be
\ddot{\tilde{\Delta}}+2\(\frac{2+\xi}{1+\xi}\)H\dot{\tilde{\Delta}}-4\pi G\rho\frac{\xi}{1+\xi}\(1+\tilde{\Delta}\)=0
\ee

Evolving this equation with the initial conditions $\tilde{\Delta}(t_i)=\dot{\tilde{\Delta}}(t_i)=0$, one obtains the almost identical results as shown in Fig. \ref{Delta_vs_z} for scales up to $.01h$ Mpc$^{-1}$. In other words, this shows that the large error in Figs. \ref{Delta_vs_k} and \ref{Delta_vs_z} are (largely) the result of ignoring the $3H^2\Phi$ term in comparison to the $\(k^2/a^2\)\Phi$ term for large scales.

\begin{table*}
	\centering
		\begin{tabular}{|c|l|}
		\hline
		\textbf{Variable} & \textbf{Definition}\\\hline
		$\delta$        & growth according to numerical integration of equations \eqref{Lper1}-\eqref{Lper3}\\\hline
		$\delta_g$      & growth according to the usual growth equation Eq.~\ref{growth} \\\hline
		$\delta_{gi}$   & growth according to the improved growth equation Eq.~\ref{growthbetter} \\\hline
		$\Delta$        & $\(\delta_g-\delta\)/\delta$ in Figs.\ref{Delta_vs_k} and \ref{Delta_vs_z}  and    $\(\delta_{gi}-\delta\)/\delta$ in Figs.\ref{Delta_vs_k_improved} and \ref{Delta_vs_z_improved}\\\hline
		$\tilde{\Delta}$ & $\(\delta_g-\delta_{gi}\)/\delta_{gi}$\\\hline
		$G$ & $d\textnormal{ln}(\delta/a)/d\textnormal{ln}a$\\\hline
		$f$ & $d\textnormal{ln}\delta/d\textnormal{ln}a$\\\hline
		\end{tabular}
	\caption{List of growth variables defined in the text.}
	\label{deltas}
\end{table*}

We wish to emphasize that contrary to what is often suggested in the literature (e.g. \cite{acquaviva}), the growth factor is not scale independent. This can clearly be seen from the form of the $\xi$ correction Eq.~\eqref{xi2}. Hence an observed scale dependence in the growth factor cannot be taken to imply a departure from Einstein gravity.

Finally, note that the other versions of the growth equation mentioned previously need to modified by such a term as well.  For example, in terms of the growth factor $G$, the growth equation Eq.(\ref{Ggrowth}) becomes
\ba
\label{Ggrowthxi}\nonumber
\frac{dG}{d\textnormal{ln}a} &+& \left(4 + \frac{1}{2}\frac{d\textnormal{ln} H^2}{d \textnormal{ln} a}\right)G + G^2\\ &+& 3 + \frac{1}{2}\frac{d \textnormal{ln} H^2}{d \textnormal{ln} a} - \frac{3}{2}\frac{\Omega_{m}(a)}{1 + \xi} = 0
\ea

Based on the above discussion, we recommend the use of Eq.~\eqref{growthbetter} over the usual Eq.~\eqref{growth} as a much better approximation to the linear growth of matter perturbations on large scales. The importance of using a version of the growth equation accurate to large scales and high redshifts is underscored by the fact that scales on the order of .01h Mpc$^-1$ and redshifts $>1$ are within the reach of future surveys such as ADEPT \cite{adept} as well as surveys based on the 21 centimeter emission. 

\section{Probing beyond-Einstein physics}
\label{Probing beyond-Einstein physics}

We now examine the implications of the failure of the usual growth equation on some recent studies which have used the growth equation as a probe to distinguish Einstein gravity from new physics. We first consider the work by Linder \cite{Linder:2005in} and Linder and Cahn \cite{lindercahn}. In \cite{Linder:2005in}, the author finds that the linear growth of structure according to Einstein gravity (represented by Eq.~\eqref{growth}) can be modeled with remarkable accuracy by a single parameter $\gamma\simeq 0.55$ for the entire expansion history of the Universe. In particular, if one uses Eq.~\eqref{Ggrowth}, then the growth history, characterized by the quantity $G(z)\equiv d\ln\(\delta/a\)/d\ln a$ can be described by the relationship
\be
\label{gammadef}
G(a)=\Omega_m\(a\)^{\gamma}-1
\ee
The parameter $\gamma$, dubbed the ``growth index'', is given by the fitting formulas
\ba
\label{linder1}\gamma&=&0.55+.02\[1+w\(z=1\)\],\qquad w<-1\\
\label{linder2}\gamma&=&0.55+.05\[1+w\(z=1\)\],\qquad w>-1
\ea
over the whole range $0<a<1$. 

In \cite{lindercahn} the authors provide analytical arguments supporting the above formulas, and contend that the narrow range in which this parameter is constrained in the context of Einstein gravity, allows for a possible distinction of Einstein gravity from other gravitational theories in which this parameter is not constrained in this range.  It is not difficult to repeat the steps of their analytic calculation, but now with the $\xi$ term included.   One can solve Eq.(\ref{Ggrowthxi}) formally 
\ba
G(a) &=& -1 + \frac{1}{a^4H(a)}\int_0^a\frac{da'}{a'}(a'^4H(a'))\\\nonumber&\times&\left(1 + \frac{3}{2(1+\xi)} -\frac{3}{2}\frac{\Omega_w}{1+\xi} - G(a')^2\right)
\ea
where the substitution $\Omega_m = 1 - \Omega_w$, where $\Omega_w$ is the dark energy fraction, has been made.  In the matter dominated regime we use the fact that $\Omega_m \gg \Omega_w$.  The $G^2$ term can be neglected, and using the form for the Hubble parameter in a matter dominated universe: $H^2 = H_0^2\Omega_m a^{-3}(1 + \Omega_w(a)/\Omega_m(a))$, $G$ is found to be
\ba
\label{Ganalytic}
G(a) &=& -1 + \frac{2}{5}\left(1 - \frac{\Omega_{w}}{2}\right) \\\nonumber&+& a^{-5/2}\left(1 - \frac{\Omega_{w}}{2}\right)\int_0^a\frac{da'}{a'}a'^{5/2}\frac{3}{2(1 + \xi)}\\\nonumber&+& a^{-5/2}\int_0^a\frac{da'}{a'}a'^{5/2}\Omega_w\left(\frac{1}{2}-\frac{3}{4(1 + \xi)}\right)
\ea
The corrected, $\xi$ dependent, growth index can then be recovered by using
\ba
\gamma &\approx& -\frac{G(a)}{\Omega_w(a)}
\ea
in Eq.(\ref{Ganalytic}))

To see how well the fit described in equations \eqref{gammadef}-\eqref{linder2} works for the true growth (as opposed to the growth predicted by the Eq.~\eqref{growth}) we plot the growth function $f\equiv d\ln\delta/d\ln a=G+1$ as derived from the true growth (on the plots this is denoted by $f(z;\delta)$ ) against the fit $\Omega_m\(a\)^{0.55}$ for the redshift range $0\leq z\leq 10$.  The growth function derived from the growth equation (denoted by $f(z;\delta_g)$ ) is shown for reference. 

\begin{figure}
	\epsfig{file=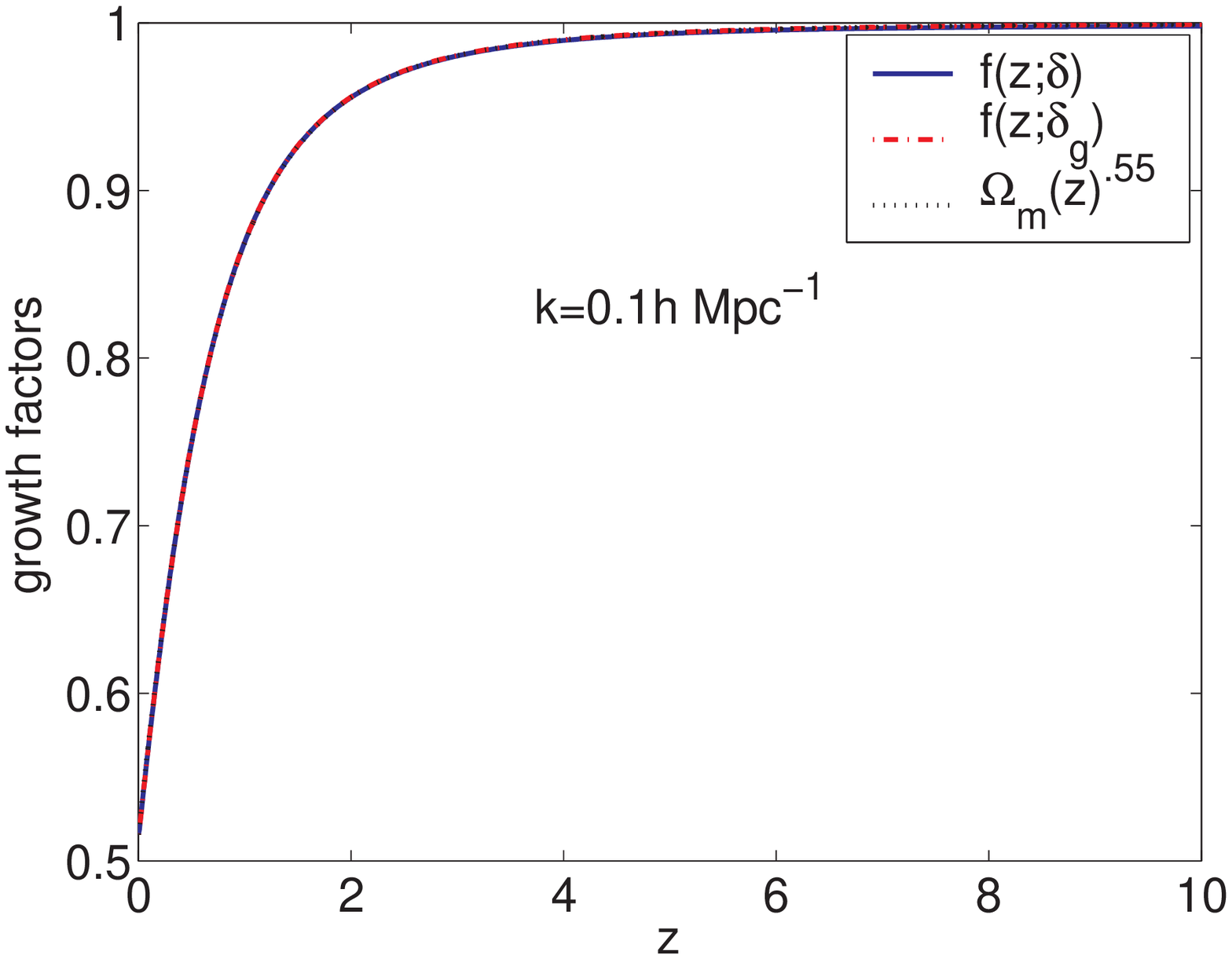,height=55mm}
	\epsfig{file=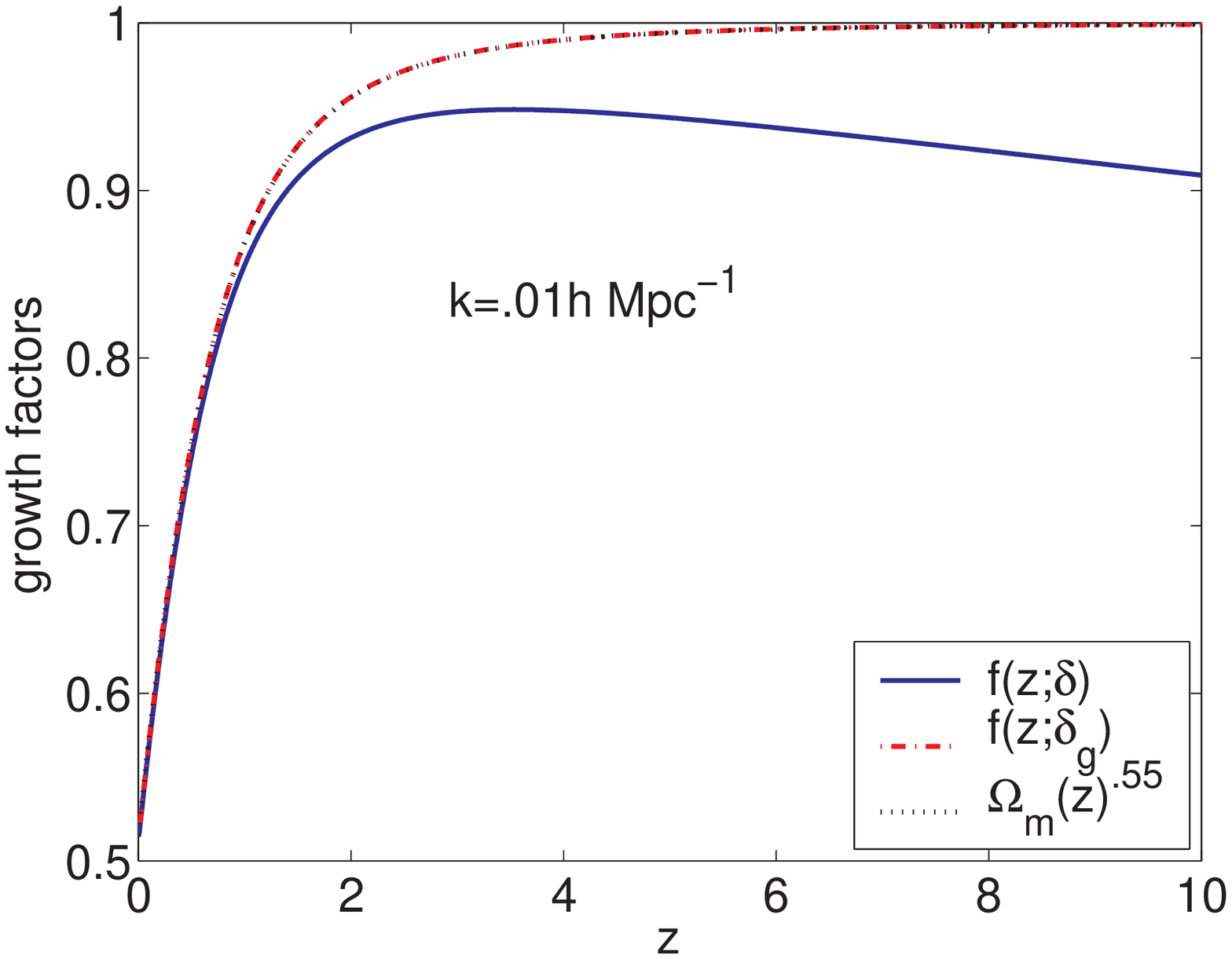,height=55mm}
	\epsfig{file=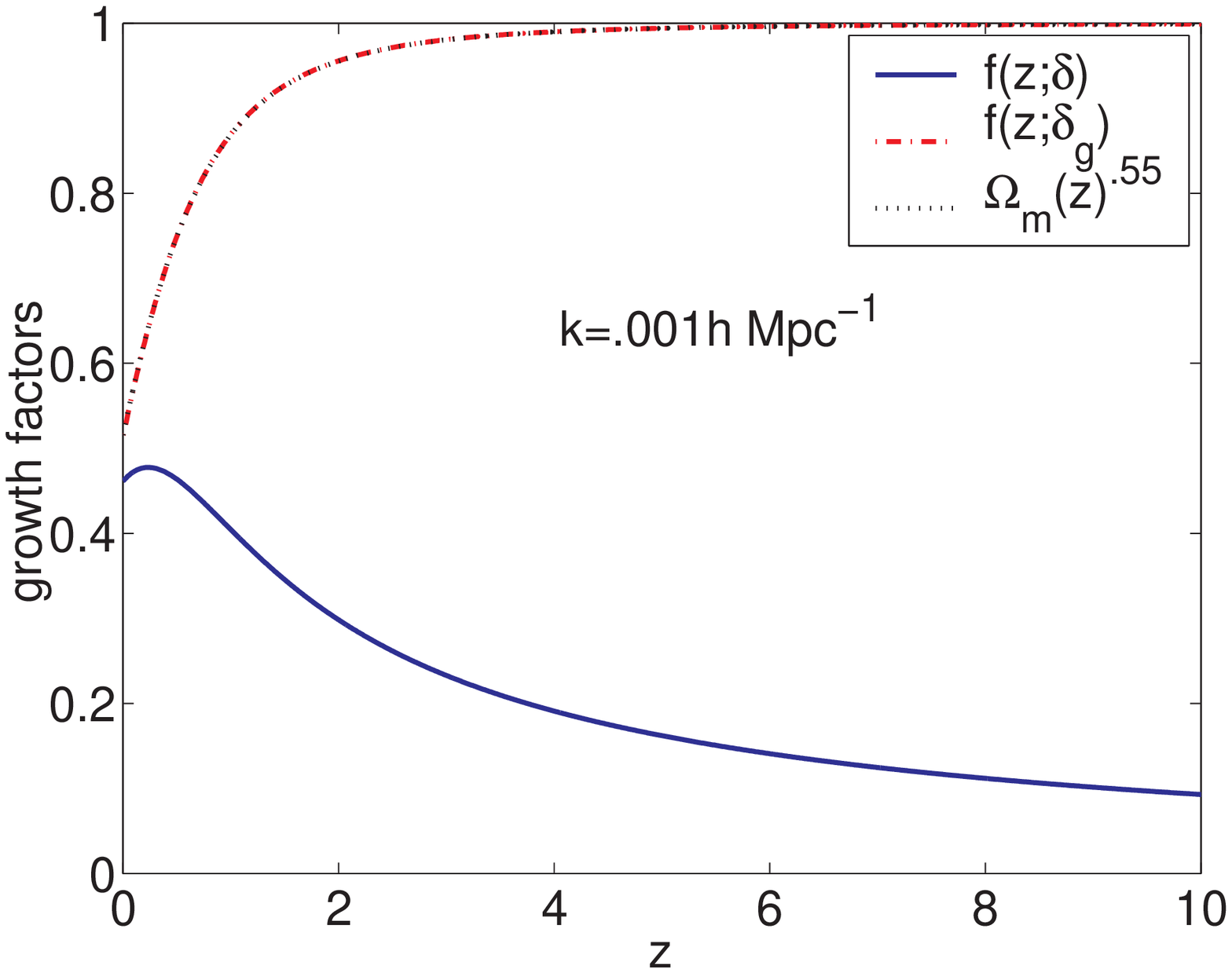,height=55mm}
	\caption
	{\label{ffits}
The growth parameter $f(z;\delta)$ derived from the true growth, plotted against the numerical fit $\Omega_{m} \(z\)^{.55}$ for three different scales. The growth parameter $f(z;\delta_g)$ derived from the growth equation is shown for reference. }
\end{figure}

Fig. \ref{ffits} confirms the findings of \cite{Linder:2005in,lindercahn} for the growth equation (and hence for small scales), but shows a drastic departure of the $\gamma$-fits from the true growth for scales $k\leq.01h$. For these large scales, it turns out that $\gamma$ has a very strong $z$-dependence, and the simple fits of Eq.~\eqref{linder1}-\eqref{linder2} do not reflect the true growth, even approximately. 

\begin{figure}
	\epsfig{file=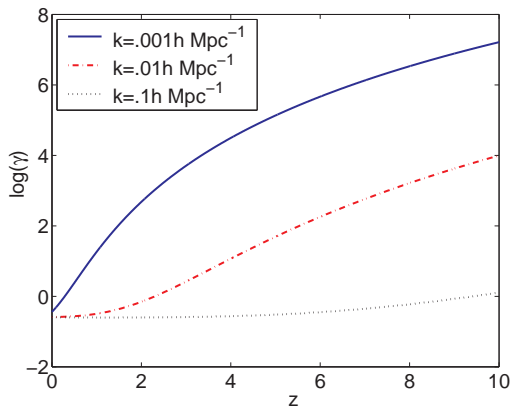,height=55mm}
	\caption
	{\label{log_gam_z}
The growth index $\gamma(z)$ derived from the true growth, plotted against redshift for three different scales. Contrary to the analysis in \cite{Linder:2005in,lindercahn}, $\gamma$ seems to have a strong and highly non-linear $z$-dependence on large scales. }
\end{figure}

For the true growth on large scales, behavior of the growth index $\gamma=\ln(f)/\ln(\Omega_m)$ is shown for three different scales $k=.1$h Mpc$^{-1}$,$k=.01$h Mpc$^{-1}$ and $k=.001$h Mpc$^{-1}$ in Fig. \ref{log_gam_z}. It is clear that for the true growth on large scales, $\gamma$ has a strongly non-linear $z$ dependence and is not by any means restricted to a small range. Hence on these scales it cannot be used to discriminate between GR and modified gravity theories.

Polarski and Gannouji \cite{polarski07} have proposed an interesting strategy for discriminating GR from modified gravity, using the present values of $\gamma_0\equiv \gamma(z=0)$ and its derivative $\gamma'_0=\gamma'(z=0)$. Starting with the growth equation in the form of Eq.~\eqref{fgrowth}, they obtain the following constraint condition linking $\gamma_0$, $\gamma'_0$, $\Omega_{m,0}$, and $w_{\rm eff,0}\equiv w_{DE,0}\Omega_{DE,0}$ (where $w_{\rm DE}$ is the equation of state parameter of the dark energy component):

\be
\label{polarski}
\begin{split}
\gamma_0' =   \left(\textnormal{ln}\Omega_{m,0}^{-1}\right)^{-1}\left[-\Omega_{m,0}^{\gamma_0}- 3(\gamma_0 -\frac{1}{2})w_{\rm eff,0}\right. \qquad\qquad\\
\left. + \frac{3}{2}\Omega_{m,0}^{1-\gamma_0}-\frac{1}{2}\right]
\end{split}
\ee 
Based on this equation, Polarski et. al.  then derive the constraint $\vert\gamma'_0\vert<0.02$ for $\Lambda$CDM. They therefore conclude that a measurement of $\gamma'_0$ outside this range would signal a departure from $\Lambda$CDM. 

However, in the light of our discussion in Sec.\ref{testing and improving}, the correct starting point should be the modified growth equation Eq.~\eqref{growthbetter}. This leads to the following modified form of the constraint equation:
\be
\label{polarski1}
\begin{split}
\gamma_0' =   \left(\textnormal{ln}\Omega_{m,0}^{-1}\right)^{-1}\left[-\Omega_{m,0}^{\gamma_0}- 3(\gamma_0 -\frac{1}{2})w_0\right. \qquad\qquad\\
\left. + \frac{3}{2}\frac{\Omega_{m,0}^{1-\gamma_0}}{1 + \xi}-\frac{1}{2}\right]
\end{split}
\ee 

Clearly, the $\xi$-correction can cause the value of $\gamma'_0$ to depart from the range observed by Polarski et. al. quite drastically. As an example, consider the parameter choices $w_{\rm DE}=-1$, $\Omega_{m,0}=0.3$ and $\gamma_0=0.554$. Eq.~\eqref{polarski} yields $\gamma'_0=-.0192$, which is within the range mentioned above. For scales of  $k\geq 0.1$h Mpc$^{-1}$ and smaller, Eq.~\eqref{polarski1} agrees. However, for $k= 0.01h$ Mpc$^{-1}$, and $k= 0.001h$ Mpc$^{-1}$, Eq.~\eqref{polarski1} yields $\gamma'_0=-0.040$ and $\gamma'_0=-0.565$ respectively, indicating that the bounds derived using the growth equation are not respected on large scales. 

\begin{figure}
	\epsfig{file=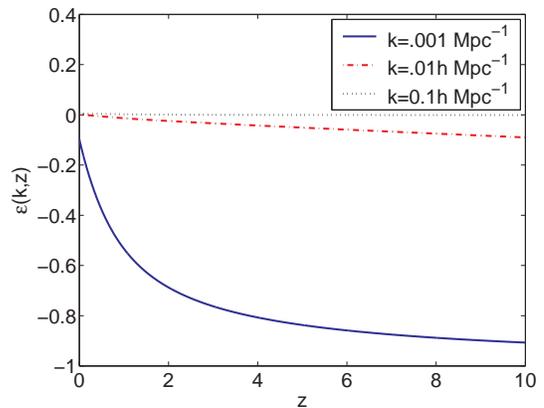,height=55mm}
	\caption
	{\label{epsilon}
The null test parameter $\epsilon(k,z)$ as proposed in \cite{acquaviva} as a function of redshift for different scales.}
\end{figure}

Acquaviva et. al \cite{acquaviva} have suggested a null test parameter $\epsilon\(k,a\)\equiv\Omega_{m}\(a\)^{-\gamma\(z\)} f\(a\)-1$ (where $\gamma(z)$ is given by Eq.~\eqref{linder2}) as a tool to discriminate GR from modified gravity . They claim that any non-zero measurement of $\epsilon$ which cannot be attributed to systematics should be interpreted as an signature of modified gravity. However, it is clear that the $\epsilon$ parameter is based on the growth equation Eq.~\eqref{growth}, and given the failure of the growth equation on large scales, this parameter is not reliable. To demonstrate this point, we plot the value of this parameter vs redshift in Fig. \ref{epsilon} for three scales. We see that on small scales ($k\simeq 0.1h$ Mpc$^{-1}$) the parameter is close to zero (as expected), but strongly deviates from zero for large scales and large redshifts. 

\section{Choice of Gauge}
\label{AppI}

In this section examine the relationship between the growth equation in the synchronous and conformal Newtonian gauges (so far we have not used the conformal time form of the line element, but we will here for easier comparison with previous works).  We rely heavily on the work of Ma and Bertschinger \cite{mabertschinger}, and any notational differences between our work and theirs will be specified for clarity.  The line element in the synchronous gauge is given by
\begin{eqnarray}
ds^2 = a^2(\tau)(-d\tau^2 + (\delta_{ij} + h_{ij})dx^idx^j)
\end{eqnarray}
The Einstein equations are written in terms of two functions, $h(\vec{k},\tau)$ (which is the trace of $h_{ij}$) and $\eta(\vec{k},\tau)$, where $\vec{k} = k\hat{k}$, that are defined through the Fourier integral of the metric perturbation $h_{ij}$ as
\begin{eqnarray}
h_{ij}(\vec{x},\tau) &=& \int d^3k e^{i\vec{k}\cdot\vec{x}}[\hat{k}_i\hat{k}_jh(\vec{k},\tau) \\\nonumber &+& (\hat{k}_i\hat{k}_j-\frac{1}{3}\delta_{ij})6\eta(\vec{k},\tau)]
\end{eqnarray}
In \cite{mabertschinger} the line element in conformal Newtonian gauge is written as
\begin{eqnarray}
ds^2 = a^2(\tau)[-(1 + 2\psi)d\tau^2 + (1 - 2\phi)dx^idx_i]
\end{eqnarray}
In the case we are examining where anisotropic stress is absent we can relate the variables $\phi$ and $\psi$ to our $\Phi$ as $\phi$ = $\psi$ = $\Phi$.

The variables in the synchronous and conformal Newtonian gauges can be shown to be related by
\begin{eqnarray}
\Phi &=& \frac{1}{2k^2}(h''(\vec{k},\tau) + 6\eta''(\vec{k},\tau) \\\nonumber &+& \frac{a'}{a}(h'(\vec{k},\tau) + 6\eta'(\vec{k},\tau)))\end{eqnarray}
where the prime denotes a derivative with respect to conformal time $\tau$, whereas a dot represents a derivative with respect to coordinate time $t$.  Using the variable $\alpha = (h' + 6\eta')/2k^2$ this relation can be written as
\begin{eqnarray}
\label{phiheta}
\Phi = a\dot{\alpha} + \dot{a}\alpha = \frac{d}{dt}(a\alpha)
\end{eqnarray}

The Einstein equations in the synchronous gauge in the absence of anisotropic stress are given by
\begin{eqnarray}
k^2\eta -\frac{1}{2}\frac{a'}{a}h' = -4\pi G_Na^2\delta\rho_s\\
k^2\eta' = 4\pi G_N a^2(\rho + p)\theta\\
h'' + 2\frac{a'}{a}h' - 2k^2\eta = -8\pi G_N a^2(3\delta p_s)\\
h'' + 6\eta'' + 2\frac{a'}{a}(h' + 6\eta') -2k^2\eta = 0
\end{eqnarray}
where $\theta$ is related to the fluid velocity by a divergence, $\theta = ik^jv_j$, and the subscript $s$ denotes the synchronous gauge (we  follow \cite{mabertschinger} here and use $c$ to denote the conformal Newtonian gauge as well).  In the simple case of a matter dominated universe one can derive the growth equation for $\delta_s \equiv \delta\rho_s/\rho$ from these equations, and one arrives at\begin{eqnarray}
\label{sgrowth}
\ddot{\delta_s} + 2H\dot{\delta_s} -4\pi G_N \delta_s = 0
\end{eqnarray}
We see that the growth equation is exact in the synchronous gauge in the case of matter domination.  In order to compare this to the case of the conformal Newtonian gauge we make use of the relation between the perturbations in the two gauges 
\begin{eqnarray}
\label{gaugediff}
\delta_s = \delta_c - \alpha \frac{\rho'}{\rho} = \delta_c +3Ha\alpha(1 + w)
\end{eqnarray}
(In the matter dominated regime one can obviously simplify this to $\delta_s = \delta_c + 3Ha\alpha$).  Inserting this relation into the growth equation in synchronous gauge, Eq.(\ref{sgrowth}), and using the relation Eq.(\ref{phiheta}) we find
\begin{eqnarray}
\ddot{\delta_c} + 2H\dot{\delta_c} -4\pi G_N\delta_c -3H^2\Phi +3H\dot{\Phi} = 0
\end{eqnarray}
Comparing this to the relation we found for $\delta$ in Eq.(\ref{deltadoubledot}) we see that they exactly match as one would expect.  In order to see this starting from Eq.(\ref{deltadoubledot}), one should use the approximations of matter domination and set $w, \dot{w}, \Theta, \dot{\Theta}$ to zero, and use Eq.(\ref{pereq4}) to substitute for $\ddot{\Phi}$.

What this shows is that if one wishes to use the exact growth equation, one should work in synchronous gauge.  The difference between the growth equation in the two gauges is given by the extra $\Phi$ and $\dot{\Phi}$ terms as we have noted, which cause a large deviation in the conformal Newtonian gauge from the solution to the usual growth equation.  In other words, the deviation from the usual growth equation in conformal Newtonian gauge is a mark of the deviation between the evolution of the matter perturbation in the synchronous and conformal Newtonian gauges.

In the process of observation, one is essentially viewing a weighted two-dimensional projection $\tilde{\delta}\(\hat{n}\)$ of a three-dimensional field $\delta\(\hat{n}\)$ . $\delta\(\hat{n}\)$ and $\tilde{\delta}\(\hat{n}\)$ are linked through a line-of-sight-integral

\be
\tilde{\delta}\(\hat{n}\)=\int_{0}^{\infty}dz W_{X}\(z\)\delta\(\hat{n}r\(z\),z\)
\ee
where $\hat{n}$ is a direction in on the sky, $r(z)$ is the comoving distance to a point at redshift $z$, and $W_{X}\(z\)$ is a window function which selects a range along the radial coordinate that contributes to the observable $\delta$. (For more details, see e.g. \cite{Zhao:2008bn})

$\delta\(\hat{n}\)$ is obviously a gauge-dependent quantity, but as we have demonstrated above, for subhorizon scales smaller than ($10^{-2}$h Mpc$^{-1}$), the gauge choice does not lead to a significant difference in $\delta$. It is only on scales close to the horizon that the two gauges can diverge significantly. In some recent work \cite{Wands:2009}, it has been shown that the Poisson equation is valid on all scales if one works in a comoving-orthogonal gauge. In our opinion, the question of which gauge works best in  modeling real astronomical measurements on large scales is important and merits further research.

\section{Conclusions}
\label{conclusions}
With the current and future ability of cosmological probes to collect data with high and ever increasing precision, one must be careful when incorporating approximations into calculations which will be compared to such data.  To this end, we have tested the familiar growth equation against direct numerical simulations of the growth of perturbations in a $\Lambda$CDM Universe and found it to be strikingly inaccurate on large ($k<0.1$h Mpc$^{-1}$) scales in the Newtonian gauge. We have traced the source of the inaccuracy to general relativistic corrections to the Poisson equation which become important at large scales and at large redshifts within the Newtonian gauge.   We proposes a modified version of the growth equation for use in the Newtonian gauge, which we show to be highly accurate on all scales up to a tenth of the horizon, for all redshifts. We have examined the implications of the failure of the growth equation on recent efforts to distinguish GR from modified gravity. finally we have discussed the growth equation in the synchronous gauge and demonstrated that the corrections to the Poisson equation correspond exactly to the difference between the overdensities in the two gauges. Our results have important implications on efforts to design null-test parameters to distinguish between models of gravity, as well as on constraining cosmological parameters based on observations from future probes.  As modifications to general relativity may begin to operate on large scales, one must take care to ensure that approximations are not masquerading as beyond Einstein physics.

\begin{acknowledgements}
The authors are grateful to Viviana Acquaviva, Edmund Bertschinger, Levon Pogosian,  Bob Scherrer, Tanmay Vachaspati, Tom Weiler, Peng-Jie Zhang and the anonymous referees for useful discussions. The authors acknowledge the hospitality of Los Alamos National Laboratory and St. John's College where part of this work was completed.  JBD was supported in part by U.S. DoE grant DE-FG05-85ER40226.
\end{acknowledgements}


\begin{thebibliography}{99}

\bibitem{zhang}
P. Zhang, M. Liguori, R. Bean, S. Dodelson 0704.1932

\bibitem{acquaviva}
  V.~Acquaviva, A.~Hajian, D.~N.~Spergel and S.~Das,
  arXiv:0803.2236 [astro-ph].
  

\bibitem{Linder:2005in}
  E.~V.~Linder,
  Phys.\ Rev.\  D {\bf 72}, 043529 (2005)
  [arXiv:astro-ph/0507263].

\bibitem{lindercahn}
  E.~V.~Linder and R.~N.~Cahn,
  Astropart.\ Phys.\  {\bf 28}, 481 (2007)
  [arXiv:astro-ph/0701317].


\bibitem{polarski07}
  D.~Polarski and R.~Gannouji,
  Phys.\ Lett.\  B {\bf 660}, 439 (2008)
  [arXiv:0710.1510 [astro-ph]].
  
\bibitem{dore08}
  Y.~S.~Song and O.~Dore,
  arXiv:0812.0002 [astro-ph].



\bibitem{copeland}
E. Copeland, M. Sami, and Shinji Tsujikawa, hep-th/0603057

\bibitem{Hwangtachyon}
J.~c.~Hwang and H.~Noh,
Phys.\ Rev.\ D {\bf 66}, 084009 (2002).



\bibitem{Hwang05}
J.~c.~Hwang and H.~Noh,
Phys.\ Rev.\ D {\bf 71}, 063536 (2005).

\bibitem{ddw}
  J.~B.~Dent, S.~Dutta and T.~J.~Weiler,
  arXiv:0806.3760 [astro-ph].
  
\bibitem{mabertschinger}
  C.~P.~Ma and E.~Bertschinger,
  Astrophys.\ J.\  {\bf 455}, 7 (1995)
  [arXiv:astro-ph/9506072].
  
\bibitem{Green:2005kf}
  A.~M.~Green, S.~Hofmann and D.~J.~Schwarz,
  JCAP {\bf 0508}, 003 (2005)
  [arXiv:astro-ph/0503387].
  

\bibitem{stochastic}
  A.~L.~B.~Ribeiro, A.~P.~A.~Andrade and P.~S.~Letelier,
  Phys.\ Rev.\  D {\bf 79}, 027302 (2009)
  [arXiv:0902.3272 [astro-ph.CO]]
  
\bibitem{gron}
  O.~Gron,
  Eur.\ J.\ Phys.\  {\bf 23}, 135 (2002)
  [arXiv:0801.0552 [astro-ph]].
  
\bibitem{liddle}
  N.~Bartolo, P.~S.~Corasaniti, A.~R.~Liddle and M.~Malquarti,
  Phys.\ Rev.\  D {\bf 70}, 043532 (2004)
  [arXiv:astro-ph/0311503].
  
\bibitem{duttamaor}
  S.~Dutta and I.~Maor,
  Phys.\ Rev.\  D {\bf 75}, 063507 (2007)
  [arXiv:gr-qc/0612027].
  

\bibitem{website}
\url{http://www.hep.vanderbilt.edu/~duttas/growtheqn.html}


  
\bibitem{adept}
\url{www.er.doe.gov/hep/files/pdfs/HEPAP_Bennett_Feb07.pdf}

\bibitem{Zhao:2008bn}
  G.~B.~Zhao, L.~Pogosian, A.~Silvestri and J.~Zylberberg,
  arXiv:0809.3791 [astro-ph].

\bibitem{Wands:2009}
  D.~Wands, A. ~Slosar,
  arXiv:0902.1084 [astro-ph.CO]







  
  

 


\end{thebibliography}
\end{document}